# 3C-SiC grown on Si by using a $Si_{1-x}Ge_x$ buffer layer

M. Zimbone[1], M. Zielinski[2], E.G. Barbagiovanni[1]*, C. Calabretta[1], F. La Via[1]

[1] CNR Institute for Microelectronics and Microsystems  95121 - Catania, Italy - Strada VIII, 5, Italy

[2] Savoie Technolac - Arche Bat.4 Allée du Lac d'Aiguebelette BP 267

73375 Le Bourget du Lac Cedex, France

Corresponding author E-mail: ericgasparo.barbagiovanni@imm.cnr.it

*Abstract*

Cubic silicon carbide (3C-SiC) is an emerging material for high power and new generation devices, but the development of high quality 3C-SiC layer still represents a scientific and technological challenge especially when grown on a Si substrate. In the present lecture, we discuss the use of a buffer layer between the epitaxial layer and the substrate in order to reduce the defectiveness and improve the overall quality of the SiC epi-film. In particular, we find that the morphology and the quality of the epi-film depends on the carbonization temperature and the concentration of Ge in close proximity of the $Si_{1-x}Ge_x$/SiC interface. Ge segregation at the interface influences the film quality, and in particular a [Ge]>12% in close proximity to the interface leads to the formation of poly-crystalls, while close to 10% induces a mirror like morphology. Moreover, by finely tuning the Ge concentration and carbonization temperature, crystal quality higher than that observed for SiC grown on bare silicon is achieved.

*Introduction*

Silicon carbide is an emerging material for high current and high power devices [i,ii,iii,iv,v], as well as, for devices working at high temperature [vi] and/or in harsh environments [vii,viii,ixx] and for new generation bio-devices[xi]. By changing the stacking order of the {111} layers, one changes the silicon carbide polytype. The only polytype that can be  grow on a silicon substrate is the cubic (3C-SiC) polytype. Development of the hexagonal polytpe, 4H-SiC, was considerable in recent years, while the development of 3C-SiC still represents a scientific and technological challenge. In particular, the growth of 3C-SiC on a silicon substrate is far from achieve the high quality need for the use in the microelectronic industry[xii].

3C-SiC and Si both have the same structure and stacking sequence along the [111] direction, but they have about a 20 % lattice mismatch [xiii]. This mismatch means that 4 layers of Si equate to 5 layers of SiC, which is the 4/5 rule. This physical constraint creates a fault every 5 layers leading to a stacking disorder and dislocations at the interface. Another very important issue regarding the growth of 3C-SiC on Si is the difference in the thermal expansion coefficients, which induces stress at the interface and causes cracks and/or macroscopic bending (and even the breakdown) of the wafer during cool down after SiC deposition. The above cited issues may be partially overcome by using compliance substrates.

Several different approaches can be used in order to create compliance substrates. It is possible to realize patterned surfaces able to expose different crystal orientations or exposing different silicon features [xiv,xv]. Other approaches imply the use of meso-structures such as pillars or whiskers. The methodology explored in the present article is the use of a buffer layer between Si and SiC that minimizes the Si and SiC lattice parameter difference and should be able to reduce residual stress in the wafer. Among the different kind of buffer layers, $Si_{1-x}Ge_x$ layer promises to be the most effective [xvi,xvii,xviii,xix,xx,xxi,xxii] . The use of this buffer layer implies that the growth conditions need to be specifically optimized considering the peculiarity of the chosen substrate. Thus, it is necessary to modify the process parameters such as temperature, pressure, ramp time for each step in the process. Indeed, SiC hetero-epitaxy on Si is a complex process that involves several steps such as etching, carbonization, growth and cool down. Each of these steps must be modified with respect to the standard process.

In the present lecture, we discussed the use of a $Si_{1-x}Ge_x$ buffer layer. We choose the $Si_{1-x}Ge_x$ composites for several reasons. First, Si and Ge are perfectly miscible and several different composites can be easily achieved. Second, regarding correcting the ratio 4/5, at growth temperature, only a small mismatch on the order of 2% remains between the 4 layers of Si and 5 layers of SiC. Third, for reducing the stress in the SiC epilayer induced by the difference in the thermal expansion coefficients.

*Material and Methods*

Samples comprised a 300 μm silicon (001) substrate with 2 μm of $Si_{1-x}Ge_x$ grown on top and a final 10nm thick Si cap. The Si capping layer was needed for the initial carbonization step. From simulations[xxiii], the value of [Ge] for an ideal lattice match at growth temperature was calculated to be around 12% and thus three germanium concentrations were used at 10%, 12% and 15% After the deposition of the $Si_{1-x}Ge_x$ layer and the capping layer, the substrates were diced into 1.5x1.5 cm² squares. The growth of the 3C-SiC films was realized using a hetero-epitaxial chemical vapor deposition (CVD) process in a horizontal hot-wall reactor[xxiv] on the 1.5 cm² substrates by NOVASIC

Silane (SiH$_4$) and ethylene (C$_2$H$_4$) were used as silicon and carbon precursors, respectively, and hydrogen (H$_2$) was used as a gas carrier. The entire deposition process constituted several steps: "etching", "carbonization" and "growth". Each of these steps were further composed of several sub-steps. The etching and carbonization steps were performed at a lower temperature of 900 °C and about 1000-1150 °C, respectively, using a hydrogen flux during etching, and hydrogen and ethylene flux during carbonization. Epilayer growth was performed for 1 hour at 1290°C, which is lower than the melting point of the highest Ge concentration. The final thickness of the 3C-SiC epilayer was estimated to be 1.16 um using Fourier-transform infrared spectroscopy (FTIR) and transmission electron microscopy (TEM) measurements for the 10% and 12% samples, while 2.15 um was observed for the 15% germanium sample. Doping (nitrogen) concentration was estimated by FTIR to be lower that 1e17 cm$^{-3}$. After carbonization the temperature was increased to the growth temperature and a silicon precursor was introduced into the chamber. After the growth step, the precursor gas flows were stopped, Ar was introduced into the chamber and the temperature was decreased to room temperature.Sample were named using the carbonization temperature and the Ge concentration of the buffer layer. Therefore, SiC grown on 12% Ge with a carbonization temperature of 1000 °C was called 1000°C,12% and so on.

Optical images were acquired using a Nanotronic NSPEC and scanning electron microscopy (SEM) images using a field emission SEM (Gemini Zeiss SUPRA™25) at a working distance of 6 mm and an electron beam of 5 keV with an in-lens detector for the backscattered electrons. TEM images were acquired with a JEOL 2010F TEM microscope in scanning mode.Photoluminescence (PL) spectra maps were collected using an HR800 integrated system by Horiba Jobin Yvon working in back-scattering configuration. A He-Cd laser was used for PL measurements with a wavelength of 325 nm and a power range from 1 to 10 mW corresponding to a power density of about 0.5 to 5 kW/cm$^2$. Coaxial optics with a dichroic mirror for the 325 nm light were used and the laser light was focused via a UV grade with a x40 objective onto the sample. The emitted light was dispersed by a 300 grooves/mm kinematic grating. Raman spectra were collected using the same HR800 integrated system used for the PL measurements. The Raman excitation sources were the He-Cd (325 nm) or a HeNe (633 nm) laser. In the case of the HeNe laser, the red light was focused and collected via a x100 objective. In both cases the scattered light was dispersed by a 1800 grooves/mm kinematic grating.

*Results*

In figure 1a, a schematic of the sample structure is shown as described in the materials and methods section. Silicon capping layer thickness is lower than the critical thickness for the formation of interfacial defects at the Si/Si$_{1-x}$Ge$_x$ interface, because the capping layer is strained and thus matches the Si$_{1-x}$Ge$_x$ lattice parameter. The strained Si capping layer with the larger lattice parameter of Si$_{1-x}$Ge$_x$ is used as a seed for the carbonization step, thus reducing the lattice mismatch between Si and SiC..

An FTIR spectra of sample 1000°C,10 % is shown in figure 1b. The spectrum consists of interference fringes in the spectral region between 1000 and 4000 cm$^{-1}$ and a peculiar peak at wavenumbers lower than 1000 cm$^{-1}$. From the fitting of the spectra (red line), it is possible to infer information about sample thickness and doping. In particular, from the maximum and minimum spectral interference position (for wavenumber higher than 1000 cm$^{-1}$) we can extract the thickness, while from the shape of the peak at 800 cm$^{-1}$ it is possible to extract the doping concentration. Table 1S.I. reports the thickness (in um) for samples grown at different carbonization temperatures and Ge concentrations. Doping concentration results are lower than 1e17 cm$^{-3}$. A low magnification cross section TEM image of the sample showing the $Si_{1-x}Ge_x$ interface and the SiC film is presented in figure 1c. Well defined interfaces are apparent in the image. Dark regions near the $Si_{1-x}Ge_x$/SiC interface are due to a large defect density, such as stacking faults, twins, dislocations and anti-phase boundaries. Away from the interface, defects reduce and the TEM image becomes brighter.

Figure 2 shows the optical microscopy images of samples 10% and 15% grown at temperatures of 1000 and 1150 °C. The 1150°C,10% sample surface is covered uniformly with small black dots and unresolved white dots, while sample 1150°C,15% has large black dots of diamteter 20-40 um with a peculiar internal structure. White dots are related to the well known silicon voids at the SiC/Si interface. Samples grown at 1000°C present very different morphologies. 1000°C,15% is quite rough and resembles a polycrystalline sample, while 1000°C,10% has a smooth and uniform surface structure. Nevertheless, unresolved white dots are still apparent in both samples at 1000°C. A close inspection of 1000°C,10% reveals the presence of vertical and horizontal stripes, probably due to extended defects. SEM images of the same samples analyzed in figure 2 are reported in figure 1S.I. Sample 1150°C,10% has small areas with relatively good epitaxial film and areas where the film is polycrystalline and rough. The sample 1150°C,15% presents similar features, but with small grain coarsening. Samples grown at lower temperature (1000°C) present very large grains for high germanium concentration (15%) and a polycrystalline nature, while a regular and smooth surface for the low (10 %) concentration is seen.

Micro raman spectroscopy analyses were conducted in order to understand how the morphology observed by SEM is related to the quality of the samples. We focus on the position of the transverse optical (TO) peak of Si at about 520 cm$^{-1}$ and on the TO and longitudinal optical plasma coupled (LOPC) peaks of SiC at 796 cm$^{-1}$ and 972 cm$^{-1}$, respectively. Spectral position and intensity of the TO peak in SiC gives information about the quality of the epitaxial film [xxv]. The TO Raman peak is forbidden for 3C-SiC grown on (001) substrates (in back scattering configuration) due to the selection rules, thus its presence is associated to defects, in particular, twins and poly-crystals. Moreover, a spectral shift of this peak is related to lattice strain caused by intrinsic or extrinsic stress. The inset of figure 3 shows the TO peak for samples carbonized at 1000°C. A negligible spectral peak shift was observed and thus a negligible stress was observed for all samples. The same results were observed for the 1050° and 1150°C samples. In figure 3, the TO intensity as a function of Ge concentration is shown for different carbonization temperatures. Note the logarithm scale on the y-axes. The samples grown at higher germanium concentration and low temperature (1000°C,15%, 1050°C,15% and 1000°C,12%) have a high TO intensity (approximatley 300-400 a.u.). This high intensity is related to the polycrystalline nature of the layer, whereby, the selection rules are broken with increasing TO intensity. For the same three samples a rough surface was observed in SEM images. On the contrary, by using higher temperature and high (15%) Ge concentration, the 1150°C,15% intensity decrease to 70 a.u. and higher wafer quality was achievable. Nevertheless, the overall quality of the sample is relatively low due to the presence of "black dots" as observed in the previous SEM images. The best results were obtained for low Ge content, in particular, for low carbonization temperature (1000°C ,10%) the lowest intensity (40 a.u.) was achieved. For the sake of clarity, we shown the TO intensity of the SiC grown on bare silicon substrate in the same figure 3: indicated as 0%. It is worth noting that the TO intensity of the samples grown on bare Si have a larger intensity (about 60 a.u) respect to the Si counterpart making the sample grown on $Si_{1-x}Ge_x$ of higher quality. Moreover, the 1000°C,10% sample shows improvement with respect to all samples fabricated with 0% [Ge].

Information about the electron mobility can be inferred by analyzing the Raman LOPC peak, indeed the spectral position and broadening are sensitive to the free electrons and to the mobility. Considering an electron density lower than 1e17 cm$^{-3}$, as measured by IR reflectivity spectra, we fit the spectra with the formula reported in [xxvi] in order to evaluate the carrier mobility. In figure 2S.I. the mobility values as obtained from the fit are reported. Whilst the mobility exhibits low values for 12 and 15 % Ge concentration, it is higher than 130 cm$^2$/V/s for the 1000°C,10% sample. This value is also higher than that observed by using a bare silicon substrate, validating the results acquired from the SEM images and the TO Raman spectra. In the same figure, in the inset, we show the LOPC spectra for the samples grown at 1000°C.

From the preceding analyses it is apparent that best results are obtained for the 1000°C,10% and the 1150°C,15% samples. Nevertheless, the overall quality of the film for 1150°C,15% is decreased due to the presence of the black dots evident in the optical microscopy and SEM image. In order to understand the nature of the black dots, we performed spatially resolved micro-Raman analysis. In figure 4a, we report, as an example, an SEM image of a black dot observed in figure 2 for the 1150°C,15% sample. The image shows a corona encasing a circle with high roughness. In figure 4b, the Raman TO peaks measured inside and outside the circle are shown and in figure 4c a spatially resolved image map of the intensity of LOPC peak (acquired in the range 790-810 cm$^{-1}$). The intensity of the TO peak is higher inside the circle indicating the presence of poly-crystals. The position of the peak has negligible spectral shift with respect to the theoretical value and thus a negligible stress was observed.

An interesting behavior connected also with the presence of dots, is found by analyzing the spectral position of the Si-Si peak (520 cm$^{-1}$) from the Raman spectra. In figure 5, we show the spectral position of the Si-Si peak (Si TO mode) as a function of the Ge concentration for different carbonization temperatures. In the same graph, the position of the peak before the growth of the SiC (named before growth) is plotted and the inset reports the spectra recorded for samples carbonized at 1000°C. It is interesting to note that sample obtained at a lower carbonization temperatures 1000°C,12%; 1000°C,15% and 1050°C,15% show a peak shift more pronounced respect to what is observed for the "before growth" samples. As an example for 1000°C,15% we observe a 502 cm$^{-1}$ peak, while the peak position of the "before growth" 15% is at 510 cm$^{-1}$ and the samples carbonized at 1150°C (green dots) follow the same trend observed for the "before growth" samples. A shift in this peak can be related to an increased Ge concentration in the surface or to the presence of stress in the Si-Ge layer [xxvii]. It is worth mentioning that samples with a higher Si TO shift (1000°C,15%; 1000°C,12%; 1050,15%) have a rough morphology (SEM images, figure 2) and a high SiC TO intensity (as reported in figure 3). Furthermore, it is a remarkable fact that a similar shift is found in the micro-Raman maps of the black dots presented in figure 4. As earlier mentioned, the position of the Si peak can be related to the germanium concentration in the sample and a shift could be interpreted assuming a Ge segregation near the SiC/Si$_{1-x}$Ge$_x$ interface. Assuming that segregation occurs, it is possible to calculate the amount of Ge segregated at the interface: it ranges from 17% at the 1000°C,12% sample interface to 26% for the 1000°C,15% sample. The intersection between the horizontal lines in figure 5 and the extrapolated dotted line represents the amount of segregated Ge.

In order to investigate further the Ge segregation at the SiC/Si$_{1-x}$Ge$_x$ interface, we performed scanning TEM analysis at the interface. The image is acquired in dark field where the image contrast is mainly related to atomic scattering intensity, i.e to the square of the atomic weight. Thus, dark zones are related to C while brighter zones are related to the presence of Si or Ge. In figure 3aS.I., we show the STEM image of the interface of sample 1000°C,15%. On the left side of this image, it is possible to observe a bright region related to Si$_{1-x}$Ge$_x$ while the darker zone on the right is related to SiC. Some features are apparent in the SiC layer: a vertical bright line 50 nm from the interface and some brighter rhomboidal areas. Additionally, an intensity increase is found near the interface. In figure 3b S.I., we report the intensity profile recorded along the black dotted line of figure 3aS.I.. At the interface (position 0 nm in figure 3b S.I.) we observed a small intensity increase as indicated by the arrow. In the same figure, it is possible to recognize a large peak 50 nm far from the interface relative to the vertical bright line of figure 3aS.I. . We performed energy-dispersive X-ray spectroscopy (EDX) measurements locally in the zones marked as 1 and 2 in figure 3aS.I., in order to understand if the brighter contrast is related to the presence of Ge or a higher amount of Si. The result is shown in figure 3cS.I.. In zone 1 (the brighter results) the Ge peak appears clearly in the spectra, while in zone 2 (the darker results) only carbon and silicon are observable. Analyses shows Ge presence in the SiC layer where the Si-Si Raman peak shift higher than expected (samples 1000°C,12%, 1000°C,15 % and 1050°C,15%). Moreover, it is worth recalling that these samples have a rough morphology and are polycrystalline.

We, also, investigated the Ge content near the interface between $Si_{1-x}Ge_x$ and SiC in samples with low Ge content (10%). A STEM image of sample 1150°C,10% is shown as an example in figure 6a. The image shows a sharp interface without the features observed in the figure 3S.I.. A brighter zone is recognizable at the interface while a darker region is observed 10 nm under the interface. In figure 6b, the intensity profile obtained by scanning inside the yellow rectangle of figure 6a is shown. The Ge content in SiC is negligible, but a brighter region near the interface (on the side of $Si_{1-x}Ge_x$) is apparent. Ge segregation at the interface is thus recognizable in a region that extends less than 10 nm from the $Si_{1-x}Ge_x$/SiC interface. It is possible to give a rough estimation of the amount of germanium content at the interface assuming that the Ge concentration far from the interface is the nominal one (10%). We assume that the contrast in the image (due to scattering) should be the sum of the Si and Ge contributions. Scattering intensity is proportional to the amount of the considered element (Si or Ge) and to the molecular weight squared. Following simple math, we obtained the equation:

$$x_{intf} = \frac{x_{bulk}}{\beta} + \frac{1-\beta}{\beta(\gamma^2 - 1)}$$

where

$$\gamma = \frac{Z_{Ge}}{Z_{Si}}$$
$$\beta = \frac{I_{bulk}}{I_{intf}}$$

where $x_{bulk}$ and $x_{intf}$ are the extrapolated concentration of Ge in $Si_{1-x}Ge_x$ bulk and at the interface. $I_{bulk}$ and $I_{intf}$ are the intensity observed in the bulk and at the interface, ZGe, and $Z_{si}$ are the molecular weights of Ge and Si. This formula allows us to estimate the amount of germanium segregated at the interface to be 14% in the 1150°C,10% sample (figure 6) and only 11% in the 1000°C,10% sample (figure 3S.I.).

**Discussion**

The above results allow us to gain more insight into the growth process. Let's first consider the effect of the Ge concentration in the $Si_{1-x}Ge_x$ layer. High nominal Ge contents (i.e. 15%) lead to the formation of poly-crystals: poly-crystal dots are found at high temperatures (figure 2, 1150°C,15%) while a full poly-crystal surface (figure 2, 1000°C,10%) was observed at low temperature. Decreasing the Ge concentration from 15% to 10%, the morphology improves irrespective of the carbonization temperature as evidenced by the reduction of the polycrystalline area and the reduction of the TO intensity. This behavior can be ascribed to the carbonization process. During carbonization, a high flux of carbon precursor reaches the surface of the $Si_{1-x}Ge_x$ substrate forming a thin and very defective layer of SiC: the carbonized layer. While Si and Ge are perfectly miscible, Si and C have SiC as the only stable phase and Ge results immiscible (solubility is on the order of $10^{16}$ ion/cm$^2$) in SiC. Thus, as

SiC is formed on the surface of the wafer, the germanium contained in the first carbonized layer is "pushed down" into the $Si_{1-x}Ge_x$ layer giving rise to the segregation observed with TEM and Raman. Segregation is more efficient as the Ge amount is increased. As segregation reaches a critical value poly-crystals come out. This phenomenon appears on the entire surface for the 1000°C,15% sample while it is confined to polycrystalline dots in the 1150°C,15% sample. The difference between these two samples is due to the dependence of the critical Ge concentration at interface with the carbonization temperature.

More complex is the effect of the carbonization temperature because of the interplay between contrasting effects that affect the amount of Ge segregated at the interface. The higher carbonization temperature enhances the specimen diffusion and it can:

1) allow the diffusion of Ge inside the $Si_{1-x}Ge_x$ layer, thus decreasing the segregation at the interface
2) enhance the diffusion of C in the carbonized layer, thus realizing a thicker carbonized layer and "pushing down" more Ge in the $Si_{1-x}Ge_x$ layer and thus increasing segregation.

The result depends on the interplay of these two contrasting effects.

Let's now discuss this effect on the 15% samples. Increasing the temperature, the sample morphology improves, indeed samples have a complete polycrystalline surface at 1000°C and only polycrystalline dots at 1150°C. This behavior can be explained considering that at low temperature (1000°C) Ge diffusion is inhibited (the diffusion coefficient of Ge in Si is $10^{-8}$ um$^2$/s[xxviii]) while high temperature (1150°C) allows Ge to diffuse (the diffusion coefficient $10^{-6}$ um$^2$/s[xxix]) into the $Si_{1-x}Ge_x$ layer reducing segregation at the interface. A reduced segregation induces an improvement in the sample quality and a reduction of the amount of polycrystals realizing only small polycrystalline dots (black dots of figure 2, sample 1150°C,15%). Once segregation occurs, the polycrystalline SiC seeds are formed at the interface and during SiC growth (at 1290°C) Ge is allowed to diffuse among SiC grains forming Ge nanocrystals, shown in figure 3S.I. The shift in the Raman peak observed in figure 5 is ascribable to the high content of Ge (and Si) in the SiC layer.

A different behavior is observed in the 10% samples. In these samples segregation is less pronounced and an opposite behavior (respect to the 15% ones) is observed: segregation occurs at higher temperature (1150°C,10%) while at low temperature (1000°C,10%) it is negligible. Indeed, Ge content, at the interface, is estimated to be 14 % or 11 % for high and low carbonization temperature. We can consider that higher temperatures enhance the diffusion of C in the carbonized layer making the carbonized layer thicker [xxx]. This effect increases the Ge concentration "pushed down" (segregated) at the interface. Sample carbonized at high temperature (1150°C,10%) have small black dots, rough surface (figure 2 and 2S.I.), higher TO intensity (>100 a.u.) and low mobility (<100 cm$^2$/V/s) than the lower temperature counterpart (1000°C,10%). Here diffusion of Ge in $Si_{1-x}Ge_x$ layer is less important being driven by a concentration gradient that is negligible in this case.

**Conclusion**

In conclusion, the effect of the SiGe buffer layer on SiC growth was investigated. Morphology and the overall film quality strongly depends on the carbonization temperature and the Ge concentration. In particular, we found that these two parameters affect the Ge concentration in a region ten nm from the interface because of a segregation process. We found that non-optimized Ge segregation gave rise to polycrystalline dots or a full polycrystalline surface. Nevertheless, good crystal quality was observed for 3C-SiC grown on a 10% $Si_{1-x}Ge_x$ buffer layer and at low carbonization temperature (1000°C). Moreover, by finely tuning the carbonization temperature and the Ge amount in the bulk, it was possible to achieve superior quality with respect to a film grown on bare silicon. This approach demonstrates the use of a buffer layer can be a real methodology to achieve high quality 3C-SiC epi-layers on silicon.

**Acknowledgements**

This work has been supported by the CHALLENGE project (HORIZON 2020 – NMBP – 720827).

**Figure caption**

Figure 1: (a) Schematic of the sample structure. It comprises a Si (001) substrate, $Si_{1-x}Ge_x$ buffer layer (2 um), Si capping layer (10 nm) and a 3C-SiC layer (1,1 um). Image is not in scale. (b) FTIR spectra of 1000°C,10% in the range between 400 and 4000 $cm^{-1}$. Fit is also shown in red. (c) Low magnification cross section TEM image of the previous sample. A well defined interface is apparent between Si and SiC. Dark regions near the interface are due to large amount of defects.

Figure 2: Example of surface morphology optical images. Ge content and carbonization temperature are shown at the border of the image.

Figure 3: Intensity of the 3C-SiC TO peak as a function of nominal Ge concentration for various carbonization temperatures. Results obtained on bare silicon are also reported. In the inset, spectra of samples carbonized at 1000°C are reported.

In figure 4: a) High magnification SEM image of a dot observed in the 1150°C,15% sample. b) Raman spectra of TO peaks measured inside and outside the dot. C) Raman map of the TO intensity peak (range 790-810 $cm^{-1}$) for the same dot observed in a).

Figure 5: Spectral position of the Si peak (TO mode) as a function of the Ge concentration for different carbonization temperatures. The 1000°C and 1050°C (black and red dots) samples shift more than what observed for the "before growth" samples while at 1150°C (green dots) samples follow the same trend observed for the "before growth" samples. In the inset, spectra recorded for samples carbonized at 1000°C are showed.

Figure 6 : (a) STEM cross section of the SiC/$Si_{1-x}Ge_x$ interface for 1150°C 10%.
(b) Intensity profile across the yellow rectangle drawn in the figure (a).

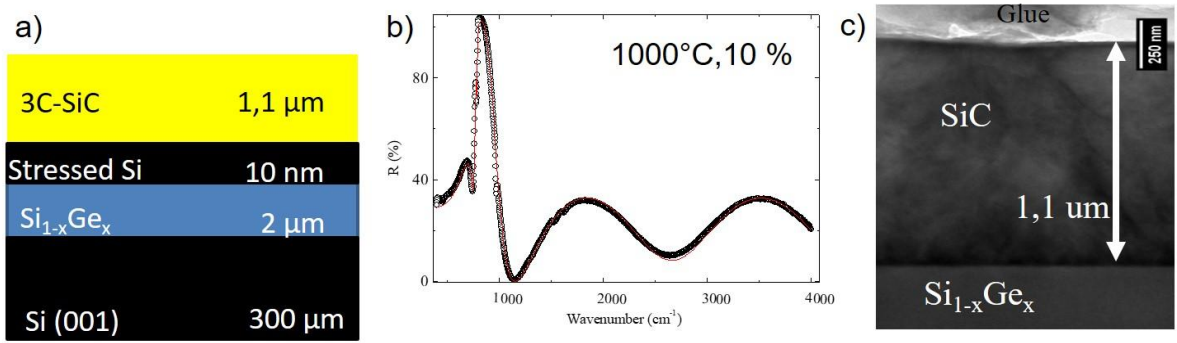

Figure 1

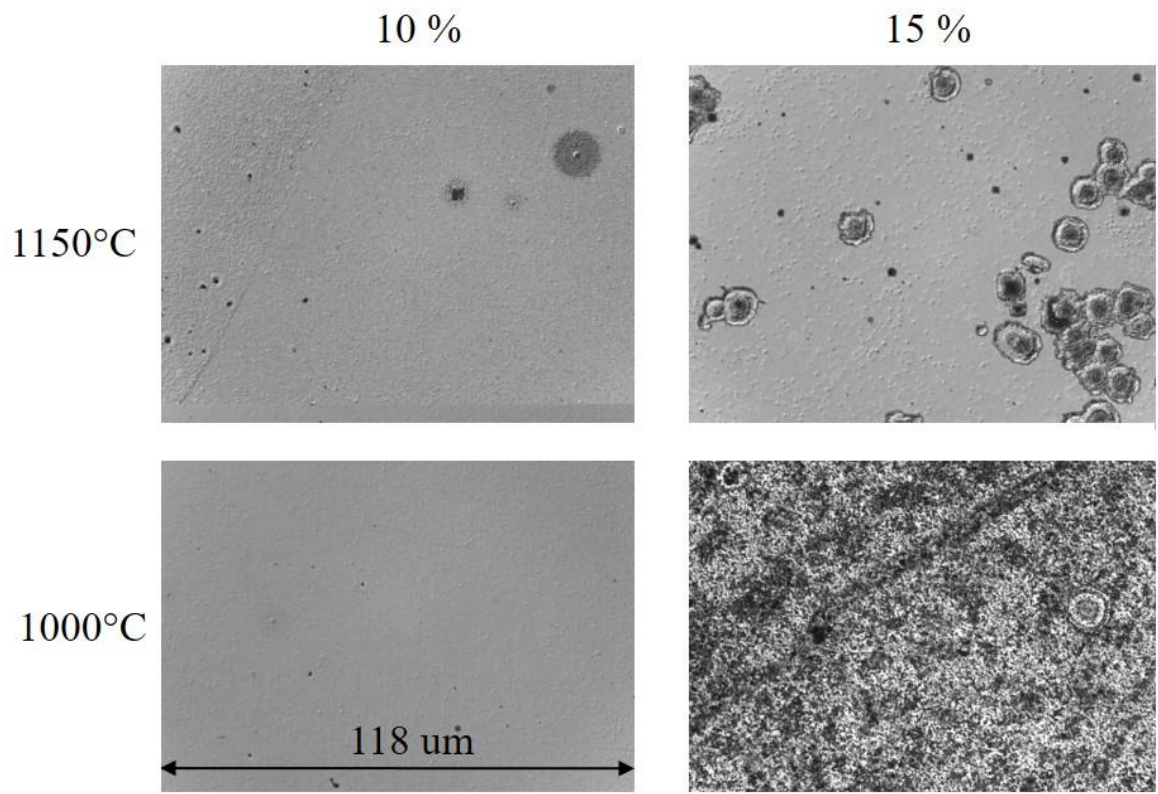

Figure 2

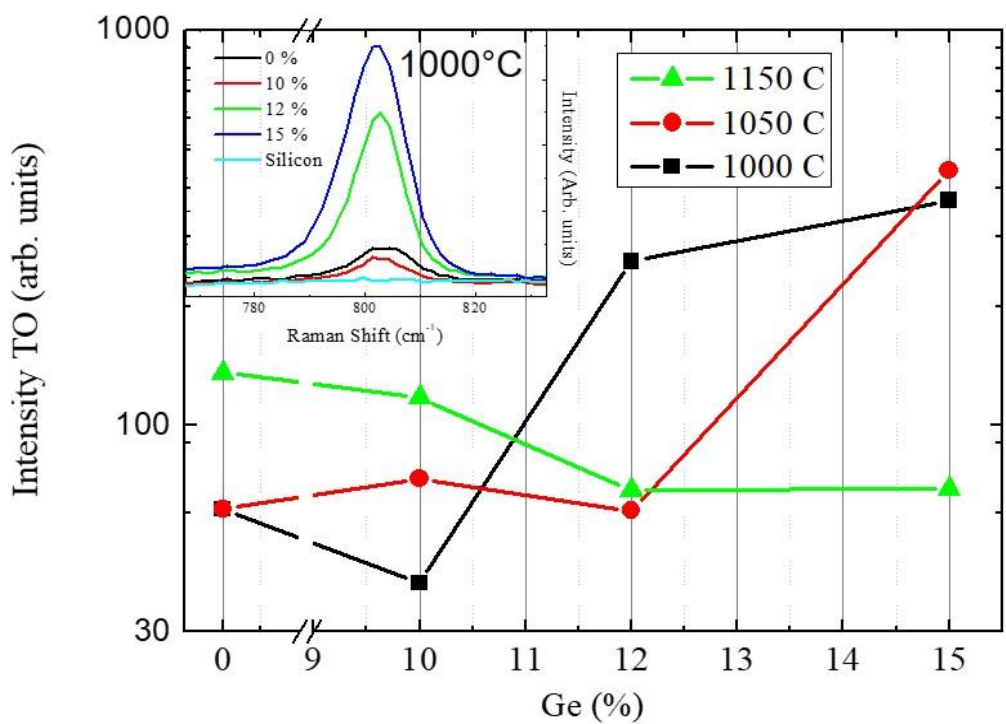

Figure 3

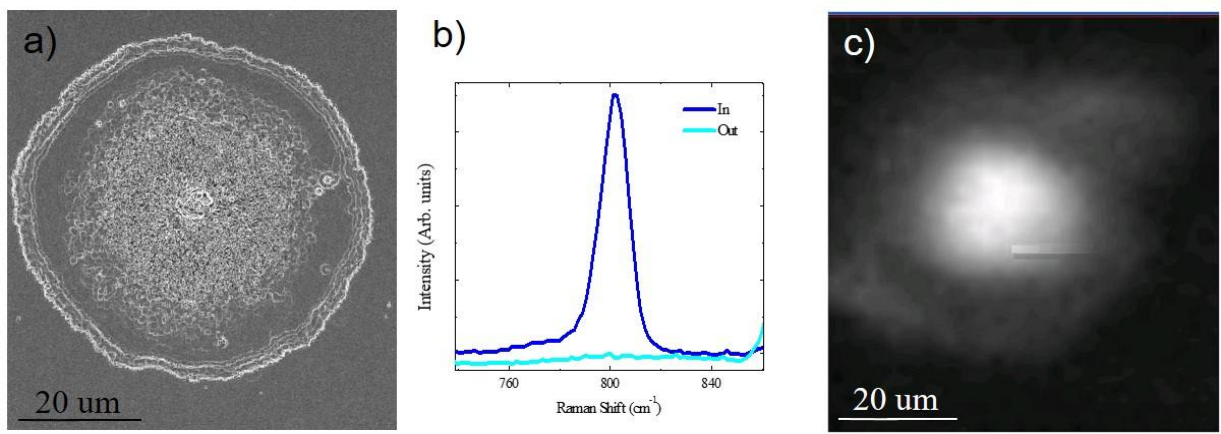

Figure 4

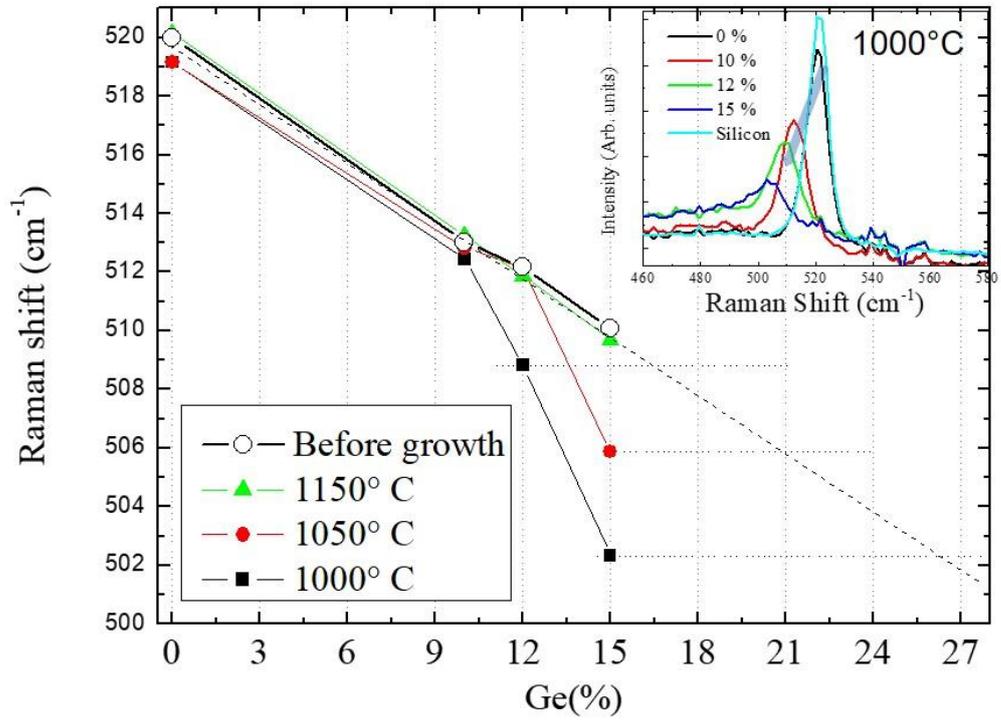

Figure 5

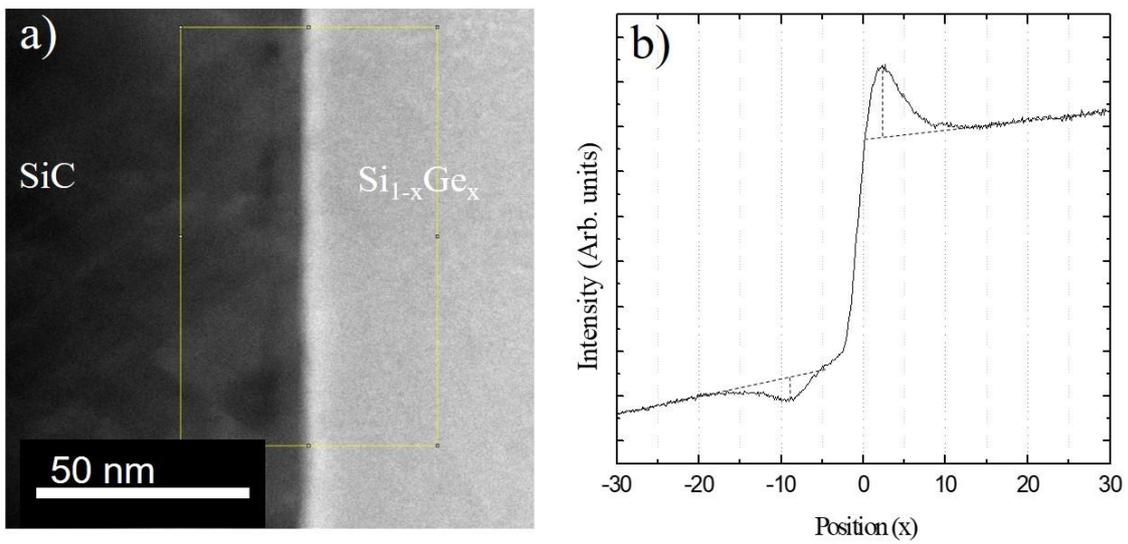

Figure 6